\documentclass[11pt]{article}
\usepackage{graphicx}
\begin{document}

\centerline{\bf\large Study on production of exotic  $0^+$ meson
$D_{sJ}^{*}(2317)$ in decays of $\psi(4415)$} \vspace{1cm}

\centerline{Xin-Heng Guo$^1$, Hong-Wei Ke$^2$, Xue-Qian Li$^2$,
Xiang Liu$^2$ and Shu-Min Zhao$^2$}

\vspace{0.8cm}

1. Institute of Low Energy Nuclear Physics, Beijing Normal
University, Beijing
100875, China\\

2. Department of Physics, Nankai University, Tianjin 300071, China

\vspace{1cm}

\centerline{\today}

\begin{abstract}
The newly observed $D_{sJ}^*$ family containing
$D_{sJ}^{*}(2317)$, $D_{sJ}(2460)$ and $D_{sJ}(2632)$ attracts
great interests. Determining their structure may be important
tasks for both theorists and experimentalists. In this work we use
the heavy quark effective theory (HQET) and a non-relativistic
model to evaluate the production rate of $D_{sJ}^{*}(2317)$ from
the decays of $\psi(4415)$, and we find that it is sizable and may
be observed at BES III and CLEO, if it is a p-wave excited state
of $D_s(1968)$. Unfortunately, the other two members of the family
cannot be  observed through decays of charmonia, because of the
constraints from the final state phase space.
\end{abstract}

\vspace{0.5cm}


\section{Introduction}

The recently observed exotic mesons $D_{sJ}^{*}(2317),\;
D_{sJ}(2460)$ and $D_{sJ}(2632)$ \cite{2317} seem to constitute a
new family of mesons which are composed of charm and strange
flavors. The mesons possess spin-parity structures of $0^+$, $1^+$
and $0^+$ respectively. This new discovery draws great interests
of both theorists and experimentalists of high energy physics.
Some authors \cite{Bardeen} suppose that $D_{sJ}^{*}(2317)$ and $
D_{sJ}(2460)$ are the chiral partners of the regular $D_s$ and
$D^*_s$, while  $D_{sJ}(2632)$ may be a radially excited state of
$D_{sJ}^*(2317)$. They may also be considered to be p-wave excited
states of $D_s$ \cite{Chao}. Alternatively, many authors suggest
that they can possiblly be  four-quark states or molecular states
\cite{four,chen}. The most peculiar phenomenon is that in some
experiments the three resonances are observed with clear signals
\cite{2317}, whereas not by other prestigious experimental groups.
One would ask if the observed resonances actually exist or the
background was misidentified as a signal. It is noted that similar
situations exist for pentaquarks \cite{lipkin}. The goal of the
research is to help designing experiments which can help
clarifying the mist.

The key point is to experimentally explore the resonances and find
a convincing explanation why they are observed in certain
experiments, but not in others. However, before it, one needs to
design certain experiments to confirm the existence  of
$D_{sJ}^{*}(2317),\; D_{sJ}(2460)$ and $D_{sJ}(2632)$ and
determine their hadronic structures. As aforementioned, there are
several different postulates. Measurements may tell which one is
more realistic.

Because of the constraint of final-state phase space, observing a
final state which involves any of the exotic states can only be
realized via decays of higher excited states in the $\psi$ family.
From the data-booklet \cite{PDG}, we can see that the lowest
excited state which can offer sufficient energy to produce
$D_{sJ}^{*}(2317)+\bar D_s(1968)$ is $\psi(4415)$, but still not
enough for $D_{sJ}^{*}(2317)+\bar D_{sJ}^{*}(2317)$. However,
since $D_{sJ}^{*}(2317)$ is a $0^+$ meson and $D_s(1968)$ is a
$0^-$ meson, a careful analysis on the total angular momentum and
parity indicates the decay of $\psi(4415)\rightarrow
D_{sJ}^{*}(2317)+\bar D_s(1968)$ is forbidden. Moreover, if only
considering the central value of $\psi(4415)$, the phase space is
not enough for $D_{sJ}^{*}(2317)+\bar D_s(1968)+\pi$ and
$D_{sJ}^{*}(2317)+\bar D_s^{*}(2112)$ which could be produced via
pure strong interaction and the only possible mode is the
radiative decay $\psi(4415)\rightarrow D_{sJ}^{*}(2317)+\bar
D_s(1968)+\gamma$. That is a decay with a three-body final state
and is an electromagnetic process where a p-wave is necessary to
conserve the total angular momentum and parity. This observation
tells us that the corresponding branching ratio must be very
suppressed and is a rare decay. Recently, Barnes et al.
\cite{Barnes} suggest to observe $D_{sJ}^{*}(2317)$ via the
process
$$\psi(4415)\rightarrow D_{sJ}^{*}(2317)+\bar D_s^{*}(2112).$$
The advantage is that it is a strong decay with a two-body final
state, therefore the amplitude may be large, but meanwhile, by the
central values, $m_{_{D_{sJ}^{*}(2317)}}+m_{_{D^*_s(2112)}}>4415$
MeV, thus this reaction can only occur via the threshold effect
and would suffer from a corresponding suppression. If it is of a
larger rate (we will estimate it later in the work), the decays
$\psi(4415)\rightarrow D_{sJ}^{*}(2317)+\bar D_s(1968)+\pi$ and
$\psi(4415)\rightarrow D_{sJ}^{*}(2317)+\bar D_s(1968)+\gamma$ can
also be realized via secondary decays of $\bar
D_s^*(2112)\rightarrow \bar D_s(1968)+\pi$ and $\bar
D_s^*(2112)\rightarrow \bar D_s(1968)+\gamma$ and these are the
dominant modes over the direct decay modes $\psi(4415)\rightarrow
D_{sJ}^{*}(2317)+\bar D_s(1968)+\pi$ and $\psi(4415)\rightarrow
D_{sJ}^{*}(2317)+\bar D_s(1968)+\gamma$ which are three-body
decays.

The picture is that the charmonium $\psi(4415)$ dissolves into a
$c\bar c$ pair and both $c$ and $\bar c$ are free and on mass
shell, and the soft gluons emitted from $c\bar c$ can excite the
physical vacuum to create a pair of $s\bar s$. The process of
$s\bar s$ pair creation is quantitatively described by the
quark-pair-creation model (QPC)\cite{Micu,qpc}. Then the $s\bar s$
join the corresponding $\bar c$ and $c$ to compose charmed mesons.
Indeed, the creation process is fully governed by the
non-perturbative QCD effects, thus the rate is not reliably
calculable so far and can only be estimated in terms of models. In
this work, we use QPC model \cite{Micu,qpc} to evaluate the rates
of $\psi(4415)\rightarrow D_{sJ}^{*}(2317)+\bar D^*_s(2112)$ and
the direct decay $\psi(4415)\rightarrow D_{sJ}^{*}(2317)+\bar
D_s(1968)+\gamma$ where a photon is emitted during the process.

In this work, we consider the transitions of
$\psi(4415)\rightarrow D_{sJ}^{*}(2317)+\bar D^*_s(2112)$ and the
subsequent observable modes $\psi(4415)\rightarrow
D_{sJ}^{*}(2317)+\bar D^*_s(2112)\rightarrow D_{sJ}^{*}(2317)+\bar
D_s(1968)+\gamma \;\;(D_{sJ}^{*}(2317)+\bar D_s(1968)+\pi)$. We
also calculate the ratio of the direct radiative decay process
$\psi(4415)\rightarrow D_{sJ}^{*}(2317)+\bar D_s(1968)+\gamma$
which is not produced via the resonance $D^*_s(2112)$.

The key point is how to evaluate the hadronic matrix elements.
Here we must adopt suitable models to do the job.

$D_{sJ}^{*}(2317)$ and $D_s^{(*)}$ all are heavy mesons, therefore
one can expect that the heavy quark effective theory (HQET)
applies for evaluating the hadronic matrix elements. For a
completeness, we keep the $1/m_c$ corrections in the formulation,
however, it is obvious that such corrections are practically
negligible in the concerned case, so that we do not really include
them in the numerical calculations. As a check, we employ a
non-relativistic model to re-evaluate the hadronic matrix elements
and compare the results obtained in the two approaches.

To obtain the concerned parameters and testify the applicability
of the model, we calculate the branching ratios of
$\psi(4040)\rightarrow D^{(*)}+\bar D^{(*)}$  and $D_s+\bar D_s$.
By fitting data, we determine the vacuum production rate of the
quark-pairs in HQET. Moreover, when using the non-relativistic
model, we also need to determine the concerned parameters in the
wavefunctions. More concretely, there are several decay channels
in $\psi(4040)$ with $c$ and $\bar c$ in the final states
($D^{0(*)}\bar D^{0(*)}$, $D^{\pm (*)}\bar D^{\mp(*)}$), and their
branching ratios are experimentally measured. Actually, in HQET,
the only free parameter is the rate of quark-pair creation from
vacuum, i.e. $\gamma_q$, then one mode is enough to fix it. We can
check the obtained model and the parameter by applying them to
evaluate other modes which have also been experimentally measured.
Our numerical results respect the pattern determined by the
experiments. For $\psi(4415)$ more channels are available, that is
$D_s\bar D_s$ (or $D_s^+D_s^-$), etc. We may naively consider that
the production of $D_s\bar D_s$ in $\psi(4415)\rightarrow D_s+\bar
D_s$ is somehow related to $\psi(4040)\rightarrow D^{(*)}+\bar
D^{(*)}$, then all the parameters  obtained from decays of
$\psi(4040)$ can be applied to study decays of $\psi(4415)$ while
assuming the parameters are not very sensitive to the energy
scale.

In both HQET and non-relativistic model, we derive the formulation
for the branching ratio of $\psi(4415)\rightarrow
D_{sJ}^{*}(2317)+\bar D^*_s(2112)$ and obtain the final numerical
results. We also formulate the direct process
$\psi(4415)\rightarrow D_{sJ}^{*}(2317)+\bar D_s(1986)+\gamma$
which is the only  channel allowed by the phase space if
neglecting the threshold effects. Even though these results with
the aforementioned approximations cannot be very accurate, one
expects that the order of magnitude of the calculated result would
be right.

If the exotic states $D_{sJ}^{*}(2317)$ is of the 4-quark
structure as suggested \cite{four,chen}, in the production process
at least three pairs of quarks are created from vacuum, and the
final state would involve more quarks and anti-quarks, thus the
integration over the final-state phase space would greatly
suppress the rate. By our rough numerical evaluation, at least 4
orders suppression would be resulted for the decays, if the exotic
meson $D_{sJ}^{*}(2317)$ is a four-quark state. Thus by measuring
the branching ratio of $\psi(4415)\rightarrow
D_{sJ}^{*}(2317)+\bar D_s^*(2112)$, we may judge (1) if the exotic
meson $D_{sJ}^{*}(2317)$ indeed exists, (2) what quark structure
it possesses.

This work is organized as follows, after the introduction, in
Sect. 2, we formulate the decay rates of $\psi(4415)\rightarrow
D_{sJ}^*(2317)+\bar{D}_{s}^*(2112)$ and direct process
$\psi(4415)\rightarrow D_{sJ}^*(2317)+\bar{D}_{s}(1968)+\gamma$.
In Sect.3, we present our numerical results along with all the
input parameters. Finally,  Sect. 4 is devoted to discussion and
conclusion. Some detailed expressions are collected in Appendix.

\section{Formulation}

The QPC model about the process that a pair of quarks with quantum
number $J^{PC}=0^{++}$ is created from vacuum was first proposed
by Micu\cite{Micu} in 1969. In the 1970s, QPC model was developed
by Yaouanc et al. \cite{qpc,yaouanc,yaouanc-1,yaouanc-book} and
applied to study hadron decays extensively. Recently there are
some works \cite{qpc-90,ackleh} to study QPC model and its
applications \cite{Zou}. In the QPC model, the interaction which
represents the mechanism of a pair of quarks created from vacuum
can be written as \cite{ackleh}
\begin{eqnarray}
\mathcal{S}_{vac}=g_{_{Iq}}\int
d^4x{\bar\psi}_q\psi_q,\;\;\;\;{\rm with}\;\;
\mathcal{L}=g_{_{Iq}}\bar\psi_q\psi_q, \label{inter}
\end{eqnarray}
where $g_{_{Iq}}=2m_{q}\gamma_{q}$. $\gamma_{q}$ is a
dimensionless constant which denotes the strength of quark pair
creation from vacuum, and can only be obtained by fitting data.
$m_q$ ($q=u,d,s$) are the masses of light quarks. In the
non-relativistic approximation \cite{yaouanc-book}, the
interaction Hamiltonian (\ref{inter}) can be expressed as
\begin{eqnarray}
\mathcal{H}_{vac}\rightarrow
\mathcal{H}^{non}_{vac}&=&\sum_{i,j}\int d \mathbf{p}_{q}
d\mathbf{p}_{\bar q}[3\gamma_{q}
\delta^{3}(\mathbf{p}_{q}+\mathbf{p}_{\bar q})\sum_{m}\langle
1,1;m, -m|0,0\rangle
\nonumber\\&&\times\mathcal{Y}_{1}^{m}(\mathbf{p}_{q}-\mathbf{p}_{\bar
q})(\chi_{1}^{-m}\varphi_{0}\omega_{0})_{i,j}]
b^{+}_{i}(\mathbf{p}_{q},s)d^{+}_{j}(\mathbf{p}_{\bar q},s'),
\end{eqnarray}
where $i$ and $j$ are SU(3)-color indices of the created quarks
and anti-quarks; $s$ and $s'$ are spin polarizations;
$\varphi_{0}=(u\bar u +d\bar d +s \bar s)/\sqrt 3$ and
$(\omega_{0})_{ij}=\delta_{ij}$ for flavor and color singlets
respectively; $\chi_{1}^{m}$ is a triplet state of spin,
$\mathcal{Y}_{1}^{m}$ is a solid harmonic polynomial corresponding
to the p-wave quark pair.

\subsection{The transition amplitude of $\psi(4415)$  in QPC model.}

\begin{figure}[htb]
\begin{center}
\begin{tabular}{ccc}
\scalebox{0.5}{\includegraphics{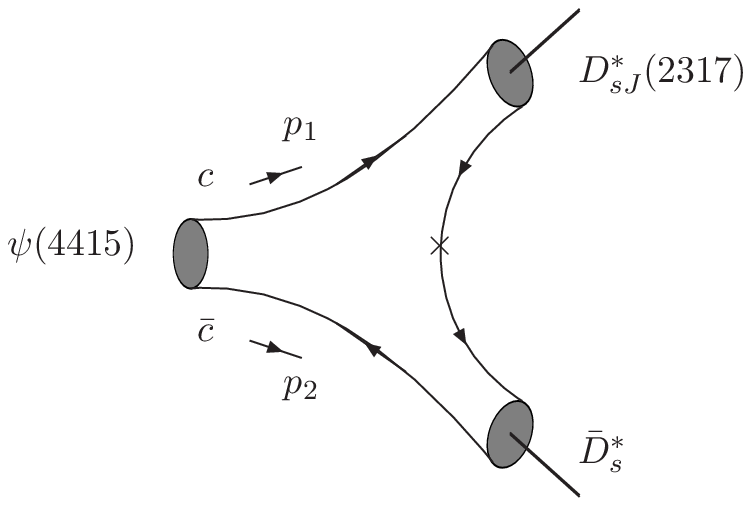}}&
\scalebox{0.5}{\includegraphics{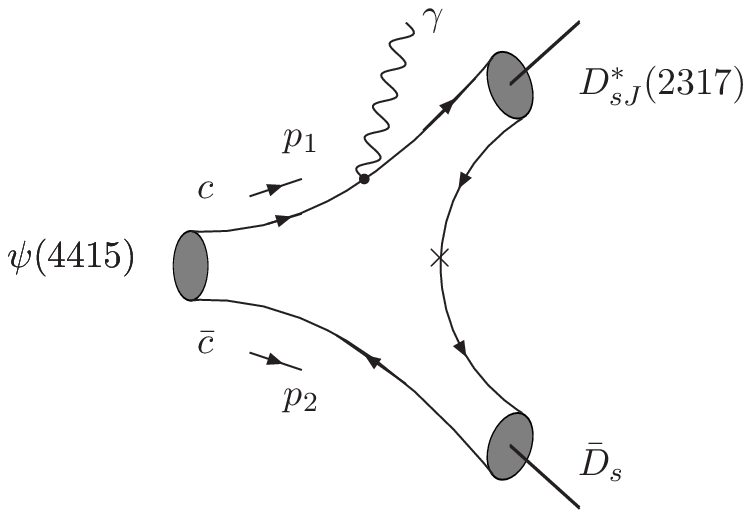}}&
\scalebox{0.5}{\includegraphics{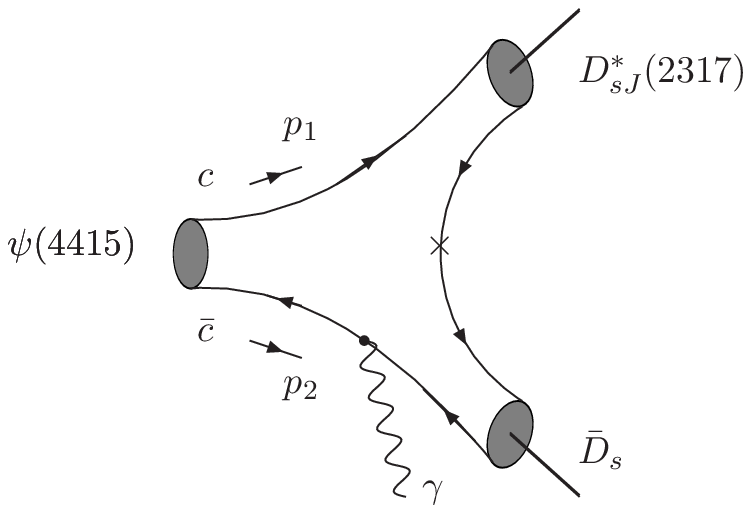}}
\\(a)&(b)& (c)\\
\scalebox{0.5}{\includegraphics{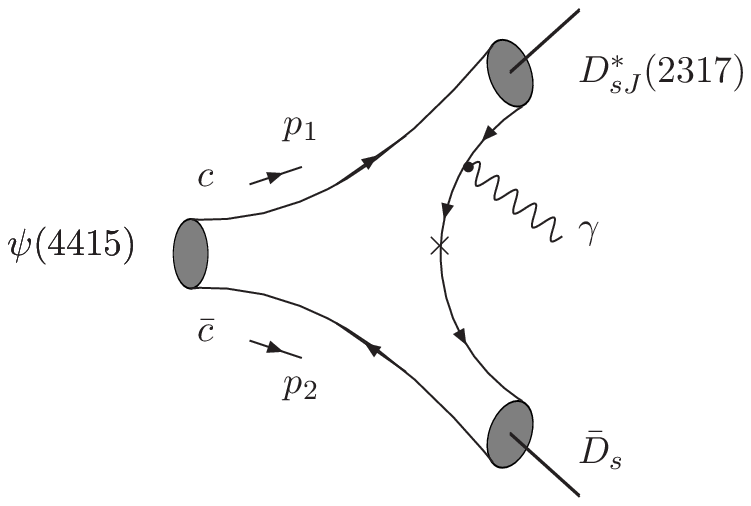}}&
\scalebox{0.5}{\includegraphics{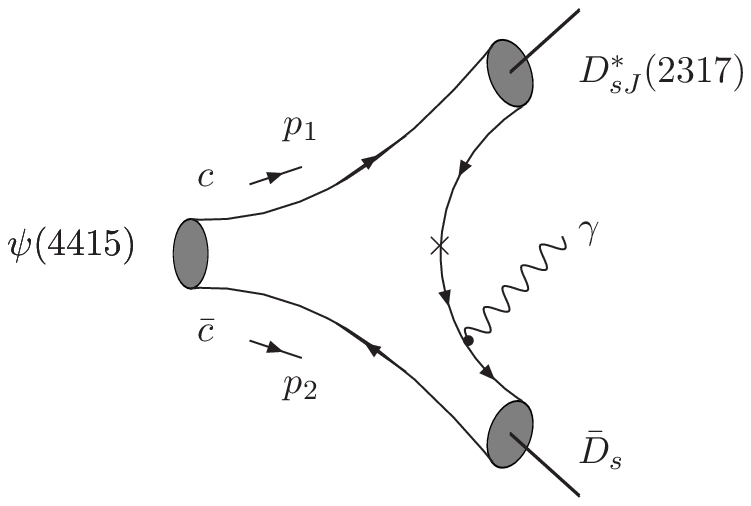}}&
\\(d)&(f)&
\end{tabular}
\end{center}
\caption{(a) is the Feynman diagram of strong decay
$\psi(4415)\rightarrow \bar{D}_{s}^{*}(2112)+D_{sJ}^*(2317)$.
(b)-(f) are the Feynman diagrams for direct $\psi(4415)\rightarrow
\gamma+\bar{D}_{s}(1968)+D_{sJ}^*(2317)$. } \label{diagram}
\end{figure}

In this work, we study the strong decay $\psi(4415)\rightarrow
\bar{D}_{s}^{*}(2112)+D_{sJ}^{*}(2317)$ and the direct radiative
decay $\psi(4415)\rightarrow \gamma
+\bar{D}_{s}(1968)+D_{sJ}^{*}(2317)$. With the QPC model, during
these transitions charm-quark and antiquark from $\psi(4415)$
combine with the $\bar ss$ created from vacuum to form final state
particles. The Feynman diagrams of these transitions are depicted
in Fig. \ref{diagram}.

The transition matrix element of $\psi(4415)\rightarrow
\bar{D}_{s}^{*}(2112)+D_{sJ}^*(2317)$ is
\begin{eqnarray}
T^{\rm{strong}}=\langle\bar{D}_{s}^{*}(2112)D_{sJ}^{*}(2317)|\mathcal{H}_{vac}(x)|\psi(4415)\rangle.
\end{eqnarray}

For the direct radiative decay  $\psi(4415)\rightarrow
\gamma+\bar{D}_{s}(1968)+D_{sJ}^*(2317)$, the transition matrix
element reads as
\begin{eqnarray}
T^{\rm{dir}}=\langle
\bar{D}_{s}(1968)D_{sJ}^{*}(2317)\gamma|\mathrm{T}\int
dxdy[\mathcal{L}_{vac}(x)\mathcal{L}_{em}(y)]|\psi(4415)\rangle,
\end{eqnarray}
where $\mathcal{L}_{em}(y)$ is the electromagnetic interaction
Hamiltonian and have the following form
\begin{eqnarray}
\mathcal{L}_{em}(y)=\pm\frac{2e}{3}\int d^{4}x\bar{\Psi}
\gamma_{\mu}\Psi A^{\mu}(y).
\end{eqnarray}
where the sign $\pm $  corresponds to charges of $c$ and $\bar c$
respectively. Considering the weak binding approximation,
$|\psi(4415)\rangle$ can be expressed as
\begin{eqnarray}
|\psi(4415)\rangle\rightarrow N\Psi(0)\bar{c}\epsilon\!\!\!\slash c|0\rangle,
\end{eqnarray}
where $\Psi(0)$ is the wave function at origin and $\epsilon$
denotes the polarization vector. $N$ is the normalization
constant.

It is also noted that for decays $\psi(4040)\rightarrow
D^{(*)}\bar D^{(*)}$, the Feynman diagrams are similar to that in
Fig.1 a.

\subsection{Evaluation of the hadronic matrix elements in HQET.}

(i) The strong decay $\psi(4415)\rightarrow
\bar{D}_{s}^{*}(2112)+D_{sJ}^{*}(2317)$.

The diagram in Fig. \ref{diagram} (a) involves the
$\mbox{q-meson-Q}$ vertices. In  refs. \cite{polosa,Hill}, the
effective Lagrangian for these vertices has been constructed based
on the heavy quark symmetry and chiral symmetry
\begin{eqnarray}
\mathcal{L}_{_{HL}}&=&\bar{h}_{v}(iv\cdot
\partial)h_{v}-[g\;\bar{\chi}(\bar{H}+\bar{S})h_{v}+H.c.]+g'[Tr(\bar{H}H)+Tr(
\bar{S}S)],\label{cl}
\end{eqnarray}
where the first term is the kinetic term of heavy quarks with
$v\!\!\!\slash h_v=h_v$; $H$ is the super-field corresponding to
the doublet $(0^{-},1^-)$ of negative parity  and has an explicit
matrix representation:
$H=\frac{1+v\!\!\!\slash}{2}(P_{\mu}^{*}\gamma^{\mu}-P\gamma_{5})$;
$P$ and $P^{*\mu}$ are the annihilation operators of pseudoscalar
and vector mesons which are normalized as $\langle
0|P|M(0^-)\rangle=\sqrt{M_{H}},\; \langle
0|P^{*\mu}|M(1^-)\rangle=\sqrt{M_{H}}\epsilon^{\mu}$; $S$ is the
super-fields related to $(0^{+},1^{+})$ and
$S=\frac{1+v\!\!\!\slash}{2}[P^{*'}_{1\mu}\gamma^{\mu}\gamma_{5}-P_{0}]$;
$\chi=\xi q$ ($q=u,d,s$ is the light quark field and
$\xi=e^{\frac{i\pi}{f}}$, here we only take the leading order as
$\xi\approx 1$).

The effective Lagrangian in Eq. (\ref{cl}) contains only the
leading order for the coupling of meson with quarks. We may also
include the $1/m_{Q}$ corrections in the expressions. The
heavy-light quark interaction Lagrangian given in \cite{Hill} is
\begin{eqnarray}
\mathcal{L}_{_{HL}}&=&\overline{Q}(i\partial\!\!\!\slash
-m_{_Q})Q-\frac{2g^2}{\Lambda^{2}}\overline{Q}\gamma_{\mu}\frac{\lambda^{A}}{2}Q\bar{\psi}
\gamma^{\mu}\frac{\lambda^{A}}{2}\psi,
\end{eqnarray}
where $Q=(b,c)$ and $\psi=(u,d,s)$, $m_{_Q}$ is the heavy quark
mass. We can obtain the $1/m_{_Q}$ corrections from two aspects.
The first comes from the quark wavefunction \cite{HQET}
\begin{eqnarray}
Q(x)=e^{-im_{_Q}v\cdot
x}\bigg(1+\frac{iD\!\!\!\slash_{\perp}}{2m_{_Q}}+\cdots\bigg)
h_{v}(x),\;\;\;D\!\!\!\slash_{\perp}=D^{\mu}-v^{\mu}v\cdot D,
\end{eqnarray} the $1/m_{_Q}$ correction is obtained by replacing $h_v$
with $(1+iD\!\!\!\slash_{\perp}/2m_{_Q})h_{v}$ in ($\ref{cl}$).
Secondly, the superfields $H$ and $S$ in (\ref{cl}) should  also
receive $1/m_{_Q}$ correction. Falk et al. \cite{falk} presented
the changes as
\begin{eqnarray*}
H\rightarrow H+\frac{1}{2m_{_Q}}[\gamma^{\mu},
iD_{\mu}H],\;\;\;S\rightarrow S+\frac{1}{2m_{_Q}}\{\gamma^{\mu},
iD_{\mu}S\}.
\end{eqnarray*}
Then, we can include the $1/m_{_Q}$ corrections in (\ref{cl}).

Now, let us write down the transition amplitude for the decay of
$\psi(4415)\rightarrow \bar{D}_{s}^{*}(2112)+D_{sJ}^{*}(2317)$ as
\begin{eqnarray}
&&\mathcal{M}(\psi(4415)\rightarrow
\bar{D}_{s}^{*}(2112)+D_{sJ}^{*}(2317))
\nonumber\\&=&\frac{\Psi(0)}{6\sqrt{M_A}}\mathrm{Tr}\bigg[
g\;\epsilon\!\!\!\slash_{_{D_{s}^{*}}}\sqrt{M_C}
\frac{i}{{p}\!\!\!\slash_{_{C}}-{p}\!\!\!\slash_{_{A}}/2-m_{s}}
g_{_{Is}}
\frac{i}{{p}\!\!\!\slash_{_{A}}/2-{p}\!\!\!\slash_{_{B}}-m_{s}}
g\sqrt{M_B}
\nonumber\\
&&\times(M_A+p\!\!\!\slash_{_{A}})\epsilon\!\!\!\slash_{A}\bigg]\nonumber\\
&=&\frac{2\Psi(0)g^2g_{_{Is}}\sqrt{M_{B}M_{C}}}{3\sqrt{M_A}}
\bigg[M_{A}m_{q}^{2}-\frac{1}{4}M_A^3 + M_{A} (p_{_A}\cdot
p_{_C})-M_{A}M_{C}^2\bigg]
(\epsilon_{_{D_{s}^{*}}}\cdot\epsilon_{A})\nonumber\\
&&\times\frac{1}{[(p_{_A}/2-p_{_B})^{2}-m_{s}^2][(p_{_A}/2-p_{_C})^{2}-m_{s}^2]},
\end{eqnarray}
where $p_{_A}$, $p_{_B}$ and $p_{_C}$ represent the four momenta
of $\psi(4415)$, $D_{sJ}^*(2317)$ and $D_{s}^{*}(2112)$; $M_{A}$
and $\epsilon_{A}$ are the mass and the polarization vector of
$\psi(4415)$; $M_{B}$ and $M_{C}$ are the masses of two produced
mesons; $g$ is the coupling constant of Q-meson-q vertex which is
given in literature\cite{polosa}. It is noted that by the central
values
$$m_{_{D^{*}_{sJ}(2317)}}+m_{_{D^*_s(2112)}}>4415\;{\rm MeV},$$ thus the
process can only occur through the threshold effect. The resonance
$\psi(4415)$ has a total width $\Gamma_A$. Considering the
distribution, we adopt the typical Gaussian form suggested by the
data group \cite{PDG}, and set the lower and upper bound for the
integration of final phase space as $M_A-\delta<M<M_A+\delta$ and
the delta-function guarantees the energy-momentum conservation.

Finally we obtain the width
\begin{eqnarray}
&&\Gamma(\psi(4415)\rightarrow
\bar{D}_{s}^{*}(2112)+D_{sJ}^{*}(2317))\nonumber\\
&=&\frac{2}{(1-\beta)\sqrt{2\pi}\;{\Gamma_A}}\int^{M_{A}+\delta}_{M_{A}-\delta}
\bigg\{\frac{1}{6M}\int
\frac{d^{3}p_{_B}d^{3}p_{_C}}{(2\pi)^{3}2E_{B}(2\pi)^{3}2E_{C}}
|\mathcal{M}(\psi(4415)\rightarrow
\bar{D}_{s}^{*}D_{sJ}^{*}(2317))
|^{2}\nonumber\\&&\times(2\pi)^{4}\delta^{4}(M-p_{_B}-p_{_C})
\bigg\}\exp\bigg[-\frac{(M-M_{A})^2}{2(\Gamma_{A}/2)^2}\bigg]dM,
\end{eqnarray}
where
\begin{eqnarray}
|\mathbf{p}_{_{B}}|&=&\frac{\sqrt{(M^2 -(M_{B}+M_{C})^2)(M^2
-(M_{B}-M_{C})^2)}}{2M},\\
E_{B}&=&\sqrt{M_{B}^{2}+\mathbf{p}_{_{B}}^2},\;\;\;
E_{C}=\sqrt{M_{C}^{2}+\mathbf{p}_{_{B}}^2};\;\;\;
\delta=1.64\frac{\Gamma_A}{2},\;\;\;\beta=10\% \cite{PDG}.
\label{defin}
\end{eqnarray}

For the indirect subsequent decays $\psi(4415)\rightarrow
D_{sJ}^*(2317)+\bar D_s^*(2112)\rightarrow D_{sJ}^*(2317)+\bar
D_s(1968)+\gamma$ and $\psi(4415)\rightarrow D_{sJ}^*(2317)+\bar
D_s^*(2112)\rightarrow D_{sJ}^*(2317)+\bar D_s(1968)+\pi$, the
rates are obtained  as
\begin{eqnarray}
&&\Gamma^{ind}(\psi(4415)\rightarrow  D_{sJ}^*(2317)+\bar
D_s(1968)+\gamma)\nonumber\\&=&\Gamma(\psi(4415)\rightarrow
D_{sJ}^*(2317)+\bar D_s^*(2112))\times BR(\bar
D_s^*(2112)\rightarrow
\bar D_s(1968)+\gamma),\\
&&\Gamma^{ind}(\psi(4415)\rightarrow  D_{sJ}^*(2317)+\bar
D_s(1968)+\pi)\nonumber\\&=&\Gamma(\psi(4415)\rightarrow
D_{sJ}^*(2317)+\bar D_s^*(2112))\times BR(\bar
D_s^*(2112)\rightarrow \bar D_s(1968)+\pi).
\end{eqnarray}

(ii) The direct radiative decay $\psi(4415)\rightarrow \gamma
+\bar{D}_{s}(1968)+D_{sJ}^*(2317)$.

This process is much more complicated than the strong decay
depicted in (i).

By the QPC model  a pair of $s\bar s$ quarks is created from
vacuum and the underlying mechanism is the soft gluon exchanges
which excite the vacuum sea. The momentum of the light quark pair
created from vacuum is small and the photon hardly has possibility
to be produced from light quark. Thus we can ignore the
contribution of Fig. \ref{diagram} (d) and (f) for the direct
$\psi(4415)\rightarrow \gamma +
\bar{D}_{s}(1968)+D_{sJ}^{*}(2317)$ process.

The amplitude for radiative decay $\psi(4415)\rightarrow \gamma+
\bar{D}_{s}(1968)+D_{sJ}^{*}(2317)$ includes several pieces. For
Fig. \ref{diagram} (b),
\begin{eqnarray}
\mathcal{M}_{(b)}&=& Q_c\sqrt{\frac{1}{m_c}}\int
\frac{\mathrm{d}\mathbf{q}}{(2\pi)^{3/2}}\psi_{_{s_{1}s_{2}}}(\mathbf{q})[\bar{v}(p_{2},s_{2})
\mathcal{O}^{(b)} u(p_{1},s_1)],\label{amplitude-1}\\
\mathcal{O}^{(b)}&=&\gamma_{5}g\sqrt{M_{C}}\frac{i}
{p\!\!\!\slash_{_{C}}-p\!\!\!\slash_{_{A}}/2-m_{s}}\cdot g_{_{Is}}
\frac{i}{-p\!\!\!\slash_{_{B}}+p\!\!\!\slash_{_{A}}/2-k\!\!\!\slash-m_{s}}
g\sqrt{M_{B}}\nonumber\\&&\times
\frac{i}{p\!\!\!\slash_{_{A}}/2-k\!\!\!\slash-m_{c}}\;\epsilon\!\!\!\slash_{k}.\nonumber
\end{eqnarray}
for Fig. \ref{diagram} (c), the transition amplitude reads as
\begin{eqnarray}
\mathcal{M}_{(c)}&=& -Q_c\sqrt{\frac{1}{m_c}}\int
\frac{\mathrm{d}\mathbf{q}}{(2\pi)^{3/2}}\psi_{_{s_{1}s_{2}}}(\mathbf{q})[\bar{v}(p_{2},s_{2})
\mathcal{O}^{(c)}u(p_{1},s_1)],\label{amplitude-2}\\
\mathcal{O}^{(c)}&=&\epsilon\!\!\!\slash_{k}
\frac{i}{k\!\!\!\slash-p\!\!\!\slash_{_{A}}/2-m_{c}}\gamma_{5}g\sqrt{M_{C}}
 \frac{i}
{p\!\!\!\slash_{_{A}}/2-p\!\!\!\slash_{_{B}}-m_{s}}
g_{_{Is}}\nonumber\\&&\times
\frac{i}{p\!\!\!\slash_{_{A}}/2-p\!\!\!\slash_{_{B}}-m_{s}}\nonumber
\;g\sqrt{M_{B}}
\end{eqnarray}
where $p_{1}$, $p_{2}$ and $s_1$, $s_2$ are the momenta and spin
projections of the charm quark and anti-charm quark, and the
following relations hold
\begin{eqnarray*}
p_{1}+p_{2}=p_{_{A}}=(M_{A},{\bf
0}),\;\;\;p_{1}-p_{2}=2q=(0,2\mathbf{q}),\;\;\;\;
\sum_{s_{1},s_{2}}\int d
\mathbf{q}|\psi_{s_{1}s_{2}}(\mathbf{q})|^2 =1.
\end{eqnarray*}
Using the method of \cite{hadronization}, we have
\begin{eqnarray}
&&\sqrt{\frac{1}{m_{c}}}\int\frac{d
\mathbf{q}}{(2\pi)^{3/2}}\psi_{s_{1}s_{2}}(\mathbf{q})\bar{v}(p_{2},s_{2})\mathcal{O}^{(i)}
u(p_{1},s_{1})\nonumber\\&&= \sqrt{\frac{1}{m_c}}\int\frac{d
\mathbf{q}}{(2\pi)^{3/2}}\psi_{s_{1}s_{2}}(\mathbf{q})Tr[(M_{A}\mathcal{O}^{(i)}_{0}
+\{\mathcal{O}^{(i)}_{0},q\!\!\!\slash\}_{+}+M_{A}q\cdot
\hat{\mathcal{O}}^{(i)})\frac{1+\gamma_{0}}{2\sqrt{2}}(-\epsilon\!\!\!\slash_{A})],
\end{eqnarray}
where
$\mathcal{O}^{(i)}_{0}=\mathcal{O}^{(i)}|_{\mathbf{q}\equiv0}$ and
$\hat{\mathcal{O}}^{(i)})\equiv\frac{\partial}{\partial
q^{\mu}}\mathcal{O}^{(i)}|_{\mathbf{q}=0}$, and $i$ denotes (b) or
(c).

Dissociation of the charmonium into $c\bar c$ can be well
described by the non-relativistic model where the wavefunction at
origin $\Psi(0)$ corresponds to the binding effect.
(\ref{amplitude-1}) and (\ref{amplitude-2}) can be further
expressed as
\begin{eqnarray}
\mathcal{M}_{(b)}&=&\frac{\Psi(0)Q_{c}}{6\sqrt{M_A}}Tr\bigg[
\gamma_{5}g\sqrt{M_{C}}\frac{i}
{p\!\!\!\slash_{_{C}}-p\!\!\!\slash_{_{A}}/2-m_{s}}\cdot
g_{_{Is}}\cdot
\frac{i}{-p\!\!\!\slash_{_{B}}+p\!\!\!\slash_{_{A}}/2-k\!\!\!\slash-m_{s}}\nonumber\\
&&\times g\sqrt{M_{B}}
\frac{i}{p\!\!\!\slash_{_{A}}/2-k\!\!\!\slash-m_{c}}\;\epsilon\!\!\!\slash_{k}
(M_{A}+p\!\!\!\slash_{_{A}})\epsilon\!\!\!\slash_{A}\bigg],\\
\mathcal{M}_{(c)}&=&-\frac{\Psi(0)Q_{c}}{6\sqrt{M_A}}Tr\bigg[
\epsilon\!\!\!\slash_{k}
\frac{i}{k\!\!\!\slash-p\!\!\!\slash_{_{A}}/2-m_{c}}\gamma_{5}g\sqrt{M_{C}}
\frac{i}
{p\!\!\!\slash_{_{A}}/2-p\!\!\!\slash_{_{B}}-m_{s}}\nonumber\\
&&\times g_{_{Is}}
\frac{i}{p\!\!\!\slash_{_{A}}/2-p\!\!\!\slash_{_{B}}-m_{s}}
\;g\sqrt{M_{B}}
(M_{A}+p\!\!\!\slash_{_{A}})\epsilon\!\!\!\slash_{A}\bigg],
\end{eqnarray}
where $p_{_{A}}$, $p_{_B}$, $p_{_C}$ and $k$ correspond to the
four momenta of $\psi(4415)$, $D_{sJ}^{*}(2317)$, $D_{s}(1968)$
and photon respectively; $\epsilon_k$ is the polarization vector
of the emitted photon.

The decay width for $\psi(4415)\rightarrow \gamma
+\bar{D}_{s}(1968)+D_{sJ}^{*}(2317)$ radiative decay is expressed
as
\begin{eqnarray}
\Gamma=\frac{1}{6M_{A}}\int\prod_{i}\bigg(\frac{d^{3}p_{_i}}{(2\pi)^{3}2E_{i}}\bigg)
(2\pi)^{4}\delta^{4}(M_{A}-p_{_{B}}-p_{_{C}}-k)|\mathcal{M}_{(b)}+\mathcal{M}_{(c)}|^{2}.\label{radial}
\end{eqnarray}
In next section, we carry out the multiple integration to obtain
numerical results.

\vspace{0.5cm}

\vspace{0.5cm}

\subsection{Evaluation of the hadronic matrix elements in the non-relativistic model.}

As discussed above, for a comparison\footnote{There is another
reason to employ the non-relativistic harmonic oscillator model.
If $D_{sJ}^*(2317)$ is of four-quark structure, the HQET  no
longer applies and the only model we can use for the
multi-constituents structure is the harmonic oscillator model.
Therefore a comparison of the results in the model with that
obtained by HQET  is indeed meaningful. Namely, HQET is believed
to be applicable in this case, thus consistence of the results
obtained in the two approaches can confirm applicability of the
harmonic oscillator model. Then we can use it to calculate the
production rate if $D_{sJ}^{*}(2317)$ is of four-quark structure.}
we are going to employ a non-relativistic model i.e. the harmonic
oscillator model to repeat the calculations made in terms of HQET.
Application of such model should be reasonable in this case.

(i) Strong decay $\psi(4415)\rightarrow
\bar{D}_{s}^{*}(2112)+D_{sJ}^{*}(2317)$.

We calculate the $\psi(4415)\rightarrow
\bar{D}_{s}^{*}(2112)+D_{sJ}^{*}(2317)$ by using QPC model in the
non-relativistic approximation. The decay width is
\begin{eqnarray}
&&\Gamma(\psi(4415)\rightarrow
\bar{D}_{s}^{*}(2112)+D_{sJ}^{*}(2317))\nonumber\\
&=&\frac{2}{(1-\beta)\sqrt{2\pi}\;{\Gamma_{A}}}\int^{M_{A}+\delta}_{M_{A}-\delta}
2\pi\frac{E_{B}E_{C}|\mathbf{k}|}{M}
\;\sum_{l,s}|\mathcal{M}_{ls}|^{2}\exp\bigg[-\frac{(M-M_{A})^2}{2(\Gamma_{A}/2)^2}\bigg]dM,
\label{strong}
\end{eqnarray}
and the concrete expression of $\sum_{l,s}|\mathcal{M}_{ls}|$ is
collected in Appendix.The definitions of $\delta$ and $\beta$ in
the expression are exactly the same as those given in
eq.(\ref{defin}).

(ii) For direct radiative decay $\psi(4415)\rightarrow \gamma
+\bar{D}_{s}(1968)+D_{sJ}^*(2317)$.

Following the traditional method \cite{yaouanc-book}, the matrix
element of radiative decay $\psi(4415)\rightarrow \gamma
+\bar{D}_{s}(1968)+D_{sJ}^*(2317)$ in nonrelativistic
approximation can be written as
\begin{eqnarray}
&&\langle
\Psi_{D_{s}}(\mathbf{p}_{2}',s_{2}';\mathbf{p}_{4}',s_{4}')
\Psi_{D_{s}(2317)}(\mathbf{p}_{1}',s_{1}';\mathbf{p}_{3}',s_{3}')\Psi_{\gamma}
(\mathbf{k},\epsilon(\mathbf{k}))
|\mathrm{T}[\mathcal{H}^{non}_{vac}\nonumber\\&&\cdot
\mathcal{H}_{em}]|\Psi_{\psi(4415)}
(\mathbf{p}_{1},s_{1};\mathbf{p}_{2},s_{2})\rangle\nonumber\\
&&=\gamma\;\mathcal{F}\;\mathcal{C}\;\sum_{n_{i}}\chi_{s_{1},s_{2}}^{n_{\psi}}
\chi_{s_{1}',s_{3}'}^{n_{_{D_{sJ}}}}\chi_{s_{2}',s_{4}'}^{n_{_{D_{s}}}}\chi_{1}^{n}
\int \prod_{a=1}^{4}d\mathbf{p}_{a}
\prod_{b=1}^{4}d\mathbf{p}_{b}'\nonumber\\&&
\times\delta^{3}(\mathbf{p}_{1}+\mathbf{p}_{2}-\mathbf{p}_{_\psi})\delta^{3}
(\mathbf{p}_{2}'+\mathbf{p}_{4}'-\mathbf{p}_{_{D_{s}}})\delta^{3}
(\mathbf{p}_{3}+\mathbf{p}_{4})\delta^{3}(\mathbf{p}_{1}'+
\mathbf{p}_{3}'-\mathbf{p}_{_{D_{sJ}}})\nonumber\\&& \times\langle
1,1;-n_{_{D_{sJ}}},n_{_{D_{sJ}}}|0,0 \rangle\langle 1,1;n,-n|0,0
\rangle \mathcal{Y}_{1}^{n}(\mathbf{p}_{3}-\mathbf{p}_{4})
\nonumber\\&&
\times\varphi_{\psi}(\mathbf{p}_{1}-\frac{1}{2}\mathbf{P}_{\psi},
\mathbf{p}_{2}-\frac{1}{2}\mathbf{p}_{_\psi})
\varphi_{_{D_{sJ}}}(\mathbf{p}_{1}'-\frac{1}{2}\mathbf{p}_{_{D_{sJ}}},
\mathbf{p}_{3}'-\frac{1}{2}\mathbf{p}_{_{D_{sJ}}})\nonumber\\&&
\times\varphi_{_{D_s}}(\mathbf{p}_{2}'-\frac{1}{2}\mathbf{p}_{_{D_s}},
\mathbf{p}_{4}'-\frac{1}{2}\mathbf{p}_{_{D_s}}) \langle
0|b_{\mathbf{p}_{1}'}d_{\mathbf{p}_{3}'}d_{\mathbf{p}_{2}'}b_{\mathbf{p}_{4}'}
a_{\mathbf{k}}\int d^{3}x
\frac{2e}{3}\bar{\Psi_{c}}\gamma^{\mu}\Psi_{c}A_{\mu}(x)
\nonumber\\&& \times
b^{\dagger}_{\mathbf{p}_{4}}d^{\dagger}_{\mathbf{p}_{3}}
d^{\dagger}_{\mathbf{p}_{2}}b^{\dagger}_{\mathbf{p}_{1}}
|0\rangle,
\end{eqnarray}
where $\mathcal{F}$ and $\mathcal{C}$ correspond to the flavor and
color factors in this transition; $\chi'$s are the spin wave
functions; $\mathbf{p}_{_\psi}$, $\mathbf{p}_{_{D_{s}}}$ and
$\mathbf{p}_{_{D_{sJ}}}$ are three-momenta  of $\psi(4415)$,
$D_{s}(1968)$ and $D_{sJ}^{*}(2317)$;  $\varphi_{\psi}$,
$\varphi_{_{D_{sJ}}}$ and $\varphi_{_{D_s}}$ are the harmonic
oscillator wave functions of $\psi(4415)$, $D_{sJ}^*(2317)$ and
$D_s(1968)$ respectively.

Neglecting some technical details, finally one can derive the
decay amplitude of Fig. \ref{diagram} (b)
\begin{eqnarray}
&&\mathcal{M}_{(b)}(\psi(4415)\rightarrow \gamma+
D^{*}_{sJ}(2317)+
\bar{D}_{s}(1968))\nonumber\\&=&i\gamma\mathcal{S}\Psi(0)
\Big(\frac{4R_{A}^3}{\sqrt{35}}\Big)\Big(\frac{R_{B}^2}{\pi}\Big)
^{{3}/{4}}\frac{\sqrt{2}}{9}
{\frac{R_{C}^{{5}/{2}}}{\pi^{{1}/{4}}}}\Big[-\frac{2e}{3}
{\Big(\frac{1}{2\pi}\Big)}^{{2}/{3}}\frac{1}{\sqrt{2E_{k}}}
(2\pi)^4\Big]\nonumber\\&&\times\int
d\mathbf{p}_{_2}\exp\Big[-\frac{1}{8}R_{A}^2(2\mathbf{p}_{_2})^2-\frac{1}{8}R_{C}^2(\mathbf{p}_{_C}+2\mathbf{p}_{_B}+2\mathbf{p}_{_2})^2
-\frac{1}{8}R_{B}^2(2\mathbf{p}_{_2}+\mathbf{p}_{_B})^2\Big]\nonumber\\&&\times\mathcal{Y}_{1}^{-n_{B}}(
-2\mathbf{p}_{_2}- \mathbf{p}_{_B}) \mathcal{Y}_{1}^{n}(
-2\mathbf{p}_{_B}-2
\mathbf{p}_{_A}+2\mathbf{p}_{_2})\bar{v}(\mathbf{p}_{_2},s_{2})\gamma_{\mu}v(\mathbf{p}_{_C}+\mathbf{p}_{_B}
-\mathbf{p}_{_A}+\mathbf{p}_{_2},s_{2'})\varepsilon_{\gamma}^\mu(\mathbf{k})
\nonumber\\&&\times\frac{-105+210 R_{A}^2\mathbf{p}_{_2}^2-84
R_{A}^4\mathbf{p}_{_2}^4 +8
R_{A}^6\mathbf{p}_{_2}^6}{12\sqrt{35}},
\end{eqnarray}
where $\mathcal{S}$ is a spin factor, $\Psi(0)$ is the wave
function of $\psi(4415)$ at origin. The indices $A$, $B$ and $C$
are for $\psi(4415)$, $D_{sJ}^{*}(2317)$ and $D_{s}(1968)$
respectively.

With the same treatment, we also obtain the amplitude
$\mathcal{M}_{(c)}$ of Fig. \ref{diagram} (c), and for saving
space we keep its expression in Appendix.

\subsection{A rough estimation of the production rate of $D_{sJ}^*(2317)$
in $\psi(4415)$ decays, if $D_{sJ}^*(2317)$ is of a four-quark
structure.}

There have been some works which suggest that $D_{sJ}^*(2317)$ is
of a four-quark structure\cite{four,chen}, the situation would be
completely different. We draw a possible Feynman diagram in Fig.2,
and one can notice that  three quark-pairs are created from
vacuum. As more particles are produced, the final state phase
space would greatly reduce the rate.

\begin{figure}[htb]
\begin{center}
\scalebox{0.7}{\includegraphics{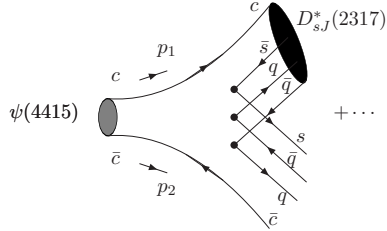}}
\end{center}
\caption{The Feynman diagram describing the production of
$D_{sJ}^{*}(2317)$ in $\psi(4415)$ inclusive decay considering the
four quark structure of $D_{sJ}^{*}(2317)$ in the QPC model, where
ellipsis denotes the diagrams for other possible quark
combinations.} \label{four}
\end{figure}

The inclusive transition matrix element can be written as
\begin{eqnarray}
\langle
\bar{c},s,q,\bar{q},D_{sJ}^{*}(2317)|T[\mathcal{H}_{vac}^{non}(x_1)\mathcal{H}_{vac}^{non}(x_2)\mathcal{H}_{vac}^{non}(x_3)
]|\psi(4415)\rangle.
\end{eqnarray}

Thus the decay width reads as
\begin{eqnarray}
&&\Gamma(\psi(4415)\rightarrow
D_{sJ}^{*}(2317)+\bar{c}+s+q+\bar{q})\nonumber\\&=&\frac{1}{6M_A}\int\frac{d^{3}p_{1}}{(2\pi)^32\omega_{1}}
\prod_{i=1}^{4}\int\frac{d^{3}k_{i}}{(2\pi)^3
}\frac{m_{i}}{E_{i}}(2\pi)^{4}\delta^{4}(M_A-p_{1}-\sum_{i=1}^{4}k_{i})\nonumber\\&&
\times|\mathcal{M}(\psi(4415)\rightarrow
D_{sJ}^{*}(2317)+\bar{c}+s+q+\bar{q})|^2.
\end{eqnarray}

Two points are noted: First this amplitude cannot be evaluated in
the framework of HQET, but only in the non-relativistic model
because of the complicated quark-structure; Secondly Fig.2 depicts
an inclusive process $\psi(4415)\rightarrow D_{sJ}^*(2317)+4\;{\rm
free\; quarks}$, considering hadronization, observable processes
can only be $\psi(4415)\rightarrow D_{sJ}^{*}(2317)+\bar
D_{s}(1968)+\pi$ and $\psi(4415)\rightarrow D_{sJ}^{*}(2317)+\bar
D_{s}(1968)+\gamma$ where $q\bar q$ annihilate into a photon. As
discussed above, such direct processes are much suppressed.

Because the inclusive decay $\psi(4415)\rightarrow D_{sJ}^*(2317)$
is related to multi-body final states, the calculation is very
complicated. The multi-integration over the phase space is very
difficult, even in terms of the Monte-Carlo method. Generally the
rate is proportional to
$$\alpha\sim (\gamma_{q})^3\bigg[\frac{4\pi}{(2\pi)^3}\bigg]^4,$$
which is a remarkable suppression factor.

Thus if $D_{sJ}^*(2317)$ is of a four-quark structure, one can
expect that the inclusive decay width of $\psi(4415)\rightarrow
D_{sJ}^*(2317)$ is at least four orders smaller than the
corresponding value  if $D_{sJ}^*(2317)$ is a p-wave excited state
of the regular $D_s$ meson.

\vspace{0.5cm}

\section{Numerical results}

(1)Determination of the concerned  parameters in the two
approaches.

(a) The parameters for the transition in HQET.

The coefficient $G=\gamma g^{2}$, which is introduced in the
transition amplitude, is obtained by fitting the decay width of
$\psi(4040)\rightarrow D \bar D$. In Appendix, we present the
formula for the decay rate of  charmonium into two charmed
pseudoscalar mesons in HQET. For calculating  $G$, the value of
$\Psi_{\psi(4040)}(0)$ is obtained by fitting the experimental
data of $\psi(4040)\rightarrow e^+ e^-$ \cite{psi(4040)}. We get
$\Psi_{\psi(4040)}(0)=0.101\;\mathrm{GeV}^{3/2}$ and
$G=12.3\;\mathrm{GeV}^{-1}$. In ref.\cite{polosa}, the value of
$g^2$ is obtained as  $g^2=4.17\; {\rm GeV}^{-1}$, thus we obtain
$\gamma_q=2.95$. Yaouanc et al. used to employ the harmonic
oscillator to evaluate such processes, and they got
$\gamma_q\approx 3$ \cite{yaouanc-book}, which is very close to
the value we obtain with HQET. However, it is also noted that
$\gamma_q$ is purely a phenomenological parameter and its value
may vary within a reasonable range, for example, in their later
work, Yaouanc et al. took $\gamma_q$ to be 4 instead, when they
fitted data\cite{yaouanc-1}.

Since there are not enough data to determine $\gamma_s$, we adopt
the relation \cite{yaouanc-1}
$$\gamma_s=\gamma_q/\sqrt 3,$$
for later numerical computations.

(b) The parameters in  the non-relativistic model.

In this scenario, the non-relativistic approximation is taken and
the expression is no longer Lorentz invariant, the relevant
parameters may be somehow different from the values in HQET,
especially the value of $\gamma_q$ which corresponds to the vacuum
creation of a quark pair. However, as pointed above the values
obtained in these two approaches are very close, thus we can use
$\gamma_q=2.95$ for later calculations.

Using the experimental results of $\psi(4040)\rightarrow
D\bar{D}$, $\psi(4040)\rightarrow D^{*}\bar{D}^{*}$ and
$\psi(4040)\rightarrow D_{s}^{+}{D}_{s}^{-}$ decays \cite{BES}
\footnote{The BES measurements of the inclusive charm cross
section at 4.03 GeV are \cite{BES}: $\sigma_{D^{0}}+
\sigma_{\bar{D}^{0}}= 19.9 \pm 0.6 \pm 2.3\; \mathrm{nb},\;
\sigma_{D^{+}}+\sigma_{\bar{D}^{-}}=6.5 \pm 0.2 \pm0.8\;
\mathrm{nb}\; {\rm and}\; \sigma_{D_{s}^{+}}+
\sigma_{D_{s}^{-}}=0.81 \pm 0.16 \pm 0.27\;\mathrm{nb}$.
Considering the relation \cite{yaouanc-book}:
$\Gamma(D^{*0}\bar{D}^{*0}):\Gamma(D^{*0}\bar{D}^{0}+D^{0}\bar{D}^{*0})
:\Gamma(D^{0}\bar{D}^{0})\approx 1:7:9$, we obtain the following
decay widths of $\psi(4040)$: $ \Gamma(\psi(4040)\rightarrow
D\bar{D})=2.97\pm
0.68\;\mathrm{MeV},\;\Gamma(\psi(4040)\rightarrow
D^{*}\bar{D^{*}})=26.73\pm 6.13\;\mathrm{MeV} \; {\rm and}\;
\Gamma(\psi(4040)\rightarrow D_{s}^{+}D_{s}^{-})=1.55\pm
0.69\;\mathrm{MeV}.$}, we obtain all the relevant parameters which
are needed for later numerical computations, For the readers'
convenience the relevant formulations \cite{yaouanc-1} are
collected in Appendix. With all the information, we obtain the
values of $R's$ in the harmonic oscillator wave functions as:
$R_{\psi}^{2}=6.00\; \mathrm{GeV}^{-2},\;R_{D}^{2}=5.25\pm 0.22
\;\mathrm{GeV}^{-2},\;\;R_{D^{*}}^{2}= 6.70 \pm 0.67\;
\mathrm{GeV}^{-2}\; {\rm and }\;R_{D_{s}}^{2}=5.20\pm 0.58\;
\mathrm{GeV}^{-2}$.

However, $R_{D_{sJ}^*(2317)}$ and $R_{D_{s}^*}$ corresponding to
$D_{sJ}^*(2317)$ and $D_{s}^*$ cannot be obtained by fitting data,
because such decay modes do not exist for the sake of phase space
of final states. As priori assumed, $D_{sJ}^*(2317)$ is a p-wave
excited state of $D_{s}$ \cite{Chao}, therefore the difference in
$R$ of $D_{sJ}^*(2317)$ and $D^{*}_{s}$ is due to the ${\bf
L}\cdot {\bf S}$ coupling which is proportional to $1/m_c$.
Furthermore, due to the heavy quark symmetry, the difference in
$R$ of $D_s^*(2112)$ and $D_{s}$ is also of order $1/m_c$, thus
$R_{D_{sJ}^*(2317)}^{2}\approx R_{D_{s}^*}^{2}\approx R_{D_{s}}^2$
are employed in the calculations of $\psi(4415)\rightarrow
\bar{D}_{s}^{*}(2112)+D_{sJ}^{*}(2317)$ and direct decay
$\psi(4415)\rightarrow \gamma
+\bar{D}_{s}(1968)+{D}_{sJ}^*(2317)$. It is believed that this
approximation is reasonable for estimating the order of magnitude
of these transitions.

(2) Our numerical results for the $D_{sJ}^*(2317)$ production.

In the calculations of $\psi(4415)\rightarrow
\bar{D}_{s}^{*}(2112)+D_{sJ}^{*}(2317)$ and the direct decay
$\psi(4415)\rightarrow \gamma+ \bar{D}_{s}(1968)+{D}_{s}^*(2317)$
by the two approaches, we employ the following parameters as
inputs: $ M_{\psi(4415)}=4.415\;\mathrm{GeV}$,
$M_{D_{s}^{\pm}}=1.968\;\mathrm{GeV}$,
$M_{D_{s}^{*}}=2.112\;\mathrm{GeV}$,
$M_{D_{sJ}^{*}(2317)}=2.317\;\mathrm{GeV}$ \cite{PDG}. By fitting
the data of $\psi(4415)\rightarrow e^+e^-$ which is available at
present, we obtain
$\Psi_{\psi(4415)}(0)=0.088\;\mathrm{GeV}^{3/2}$.

We now present the numerical results obtained with the two
approaches in Table. 2.
\begin{table}[htb]
\begin{center}
\begin{tabular}{|c|c|c|} \hline
& I & II \\
\hline $Br(\psi(4415)\rightarrow
\bar{D}_{s}^{*}(2112)+D_{sJ}^{*}(2317))$& $9.16\%$ & $9.58\%$\\
\hline $Br(\psi(4415)\rightarrow \gamma+
\bar{D}_{s}(1968)+{D}_{sJ}^*(2317))$(ind)&$8.63\%$&$9.03\%$
\\
\hline $Br(\psi(4415)\rightarrow \pi+
\bar{D}_{s}(1968)+{D}_{sJ}^*(2317))$(ind)&$5.31\times10^{-3}$&$5.56\times10^{-3}$
\\ \hline
$Br(\psi(4415)\rightarrow \gamma+
\bar{D}_{s}(1968)+{D}_{sJ}^*(2317))$(dir)&$8.46\times10^{-5}$&$2.29\times10^{-5}$
\\ \hline
\end{tabular}
\end{center}
\caption{Columns I and II correspond to the numerical results
obtained in HQET  and the non-relativistic model respectively.}
\end{table}

\vspace{0.5cm}

\section{Discussion and conclusion}

It is obvious that the newly discovered $D_{sJ}$ family may be
very significant for better understanding of the hadronic
structure and low energy QCD. The members of the family,
$D_{sJ}^*(2317),\;D_{sJ}(2460),\; D_{sJ}(2632)$, all have positive
parity, so that they cannot fit in an s-wave $c\bar s(\bar cs)$
structure. The literatures suggest that they may be p-wave excited
states, namely chiral partners of $D_s,\;D_s^*$ etc. or four-quark
states as well as molecular states. It is necessary to look for a
more plausible way to determine their configurations, i.e. design
an experiment(s) to clarify the picture. At least we would like to
find an experiment to judge (1) if such states indeed exist, (2)
their quark configuration (p-wave excited states or four-quark
states).

To have a larger production rate, it is reasonable to look for
$D_{sJ}^*(2317)$ via strong decays of higher excited states of
charmonia. The most possibly available charmonium is $\psi(4415)$.
Since $D_{sJ}^*(2317)$ mesons have positive parity, the decay mode
of $\psi(4415)\rightarrow D_{sJ}^*(2317)+\bar D_s(1968)$ is
forbidden and if considering the central values of the masses of
the concerned particles and constraints from phase space of final
states, only $\psi(4415)\rightarrow D_{sJ}^*(2317)+\bar
D_s(1968)+\gamma$ is allowed. This direct radiative decay must be
much suppressed as discussed in the introduction.

Barnes et al. \cite{Barnes} suggested to observe  decay
$\psi(4415)\rightarrow D_{sJ}^*(2317)+\bar D_s^*(2112)$ which can
occur via the threshold effects. The consequent decays
$\psi(4415)\rightarrow D_{sJ}^*(2317)+\bar D_s^*(2112)\rightarrow
D_{sJ}^*(2317)+\bar D_s(1968)+\gamma$ and $\psi(4415)\rightarrow
D_{sJ}^*(2317)+\bar D_s^*(2112)\rightarrow D_{sJ}^*(2317)+\bar
D_s(1968)+\pi$ can be observed. Even though such processes may
only occur via threshold effects and should be suppressed, it is
noted that $m_{_{D_{sJ}^*(2317)}}+m_{_{D_s^*(2112)}}$ is only
slightly above 4415 MeV, one can expect that the suppression is
not very strong.

In this work, we carefully study the production of
$D_{sJ}^*(2317)$ in the decays of $\psi(4415)$ and evaluate its
production rate. The processes are realized as the charmonium
$\psi(4415)$ dissolves into a $c\bar c$ pair which then combines
with $\bar s s$ created from vacuum due to the non-perturbative
QCD effects and constitute two mesons. The first step is
determined by the wavefunction of $\psi(4415)$ at origin and the
light-quark-pair creation is described by the QPC model
\cite{yaouanc}. For evaluating the hadronic transition matrix
elements, we employ two approaches, i.e.  HQET and the
non-relativistic model. Our final numerical results achieved in
the two approaches confirm this allegation as they are reasonably
consistent with each other.

To guarantee the plausibility of the results, we obtain all
necessary parameters by fitting data. However, it is understood
that there must be some errors from both theoretical and
experimental aspects, and the parameters should have some
uncertainties, especially the vacuum creation rate of the light
quark pair. Thus the real rates may be within a range around the
values estimated with the input parameters and theoretical
approaches, the order of magnitude should be correct and
trustworthy.

For a comparison, we have also evaluated the transition rate of
the direct radiative decay $\psi(4415)\rightarrow
D_{sJ}^*(2317)+\bar D_s(1968)+\gamma$ in the same approaches and
find that the resultant rate is two orders smaller than that
through the intermediate state $D_s^*(2112)$, even though it is
realized via the threshold effects.

We find that even though the threshold effects suppress the
production rate of $\psi(4415)\rightarrow D_{sJ}^*(2317)+\bar
D_s^*(2112)$, it is still sizable if $D_{sJ}^*(2317)$ is a p-wave
excited state. If so, it can be observed in the future experiments
of BES III, CLEO and maybe at Babar or even LHC-b. However, as our
calculations indicate that if $D_{sJ}^*(2317)$ is of a four-quark
structure, its production rate is much more suppressed and cannot
be observed in decays of charmonia.

Unfortunately, in such decays, one can only expect to observe
$D_{sJ}^*(2317)$, but not the two other members of the new family.
However, once the existence and structure of $D_{sJ}^*(2317)$ are
definitely confirmed, we have reasons to believe existence of the
other two. Moreover, we can have more knowledge on the hadronic
structure and may design experiments to testify the other two. We
are looking forward to new experimental results to clarify this
theoretical problem. In a recent work, some authors \cite{Zhu}
calculate the decay rates of $D_{sJ}^{*}(2317)$ and
$D_{sJ}(2460)$. They claim that their results prefer the ordinary
$c\bar s(\bar cs)$ quark-structure for the mesons. However, a
decisive conclusion must be drawn from a deterministic
experiment(s), and $D_{sJ}^*(2317)+\bar D_s^*(2112)$ suggested by
Barnes et al. as well as subsequent observable modes
$D_{sJ}^*(2317)+\bar D_s(1968)+\gamma$, $D_{sJ}^*(2317)+\bar
D_s(1968)+\pi$ would provide an ideal possibility to make this
judgement.

\vspace{0.6cm}

\noindent Acknowledgment:

This work is  supported by the National Natural Science Foundation
of China (NNSFC).

\vspace{0.5cm}

{\bf{Appendix}}

(a) In the Eq. (\ref{strong}), the $\sum_{l,s}|\mathcal{M}_{ls}|$
is
\begin{eqnarray}
&&\sum_{l,s}|\mathcal{M}_{ls}|\nonumber\\&=&\frac{R_{A}^{3/2}R_{B}^{3/2}R_{C}^{5/2}}
{147456\sqrt{35}\pi\eta^{11/2}}\exp\bigg[ -\frac{\mathbf{k}^2
R_{A}^{2}(R_{B}^{2}+R_{C}^2)}{8(R_{A}^{2}+R_{B}^{2}+R_{C}^{2})}\bigg]
\bigg\{R_{A}^{6}\Big[-48
a^{5}(2\zeta+1)\eta^{3}\mathbf{k}^{6}\nonumber\\&&
+8a^{6}\eta^{3}\Big(2\zeta(\zeta+1)\eta
\mathbf{k}^{2}+3\Big)\mathbf{k}^{6}-336a^{3}(2\zeta+1)\eta^{2}\mathbf{k}^{4}
+12a^{4}\eta^{2}\Big(
14\zeta(\zeta+1)\eta\mathbf{k}^{2}\nonumber\\&&+8\sqrt{3}+3\Big)\mathbf{k}^{4}
-420a(2\zeta+1)\eta\mathbf{k}^{2}+42a^{2}\eta\Big(10\zeta(\zeta+1)\eta\mathbf{k}^{2}
+8\sqrt{3}+3\Big)\mathbf{k}^{2}\nonumber\\&&
+105\Big(2\zeta(\zeta+1)\eta\mathbf{k}^{2}+9 \Big)\Big]-84\eta
R_{A}^{4}\Big[-16a^{3}(2\zeta+1)\eta^{2}\mathbf{k}^{4}+4a^{4}\eta^{2}
\Big(2\zeta(\zeta+1)\eta\mathbf{k}^{2}+3\Big)\mathbf{k}^{4}\nonumber\\&&
-40a(2\zeta+1)\eta
\mathbf{k}^{2}-4a^{2}\eta\Big(2\zeta(\zeta+1)\eta\mathbf{k}^{2}-4\sqrt{3}+21
\Big)\mathbf{k}^{2}+15\Big(2\zeta(\zeta+1)\eta\mathbf{k}^{2}+7\Big)
\Big]\nonumber\\&&
+1680\eta^{2}R_{A}^{2}\Big[6\zeta(\zeta+1)\eta\mathbf{k}^{2}-4a(2\zeta+1)
\eta\mathbf{k}^{2}+2a^{2}\eta\Big(2\zeta(2\zeta+1)\eta\mathbf{k}^{2}+3\Big)
\mathbf{k}^{2}+15\Big]\nonumber\\&&
-6720\eta^{3}\Big(2\zeta(\zeta+1)\eta\mathbf{k}^{2}+3\Big)\bigg\},
\end{eqnarray}
where
\begin{eqnarray*}
\zeta&=&\frac{R_{A}^{2}}{R_{A}^{2}+R_{B}^{2}+R_{C}^{2}},\;\;\;
\eta=\frac{R_{A}^{2}+R_{B}^{2}+R_{C}^{2}}{8},\;\;\;
a=1+\zeta.\;\;\;
\end{eqnarray*}

(b) The concrete expression of $\mathcal{M}_{_{(c)}}$ is
\begin{eqnarray}
&&\mathcal{M}_{_{(c)}}(\psi(4415)\rightarrow \gamma
+D^{*}_{s}(2317)+\bar{D}_{s}(1968))\nonumber\\&=&i\gamma\mathcal{S}\Psi(0)
\Big(\frac{4R_{A}^3}{\sqrt{35}}\Big)\Big(\frac{R_{B}^2}{\pi}\Big)
^{{3}/{4}}\frac{\sqrt{2}}{9}
{\frac{R_{C}^{{5}/{2}}}{\pi^{{1}/{4}}}}\Big[-\frac{2e}{3}
{\Big(\frac{1}{2\pi}\Big)}^{{2}/{3}}\frac{1}{\sqrt{2E_{k}}}
(2\pi)^4\Big]\nonumber\\&&\times \int
d\mathbf{p}_{_1}\exp\Big[-\frac{1}{8}R_{A}^2(2\mathbf{p}_{_1})^2-\frac{1}{8}R_{B}^2
(\mathbf{p}_{_B}+2\mathbf{p}_{_C}+2\mathbf{p}_{_1})^2
-\frac{1}{8}R_{C}^2(2\mathbf{p}_{_1}+\mathbf{p}_{_C})^2\Big]\nonumber\\&&\times
\mathcal{Y}_{1}^{-n_{B}}(
\mathbf{p}_{_B}+2\mathbf{p}_{_C}+2\mathbf{p}_{_1} )
\mathcal{Y}_{1}^{n}( -2\mathbf{p}_{_C}-2
\mathbf{p}_{_1})\bar{u}(\mathbf{p}_{_B}+\mathbf{p}_{_C}+\mathbf{p}_{_1},s_{2})\gamma_{\mu}
u(\mathbf{p}_{_1},s_{2'})\varepsilon_{\gamma}^\mu(\mathbf{k})
\nonumber\\&&\times\frac{-105+210 {R}_{A}^2 \mathbf{p}_{_2}^2-84
{R}_{A}^4\mathbf{p}_{_2}^4 +8
{R}_{A}^6\mathbf{p}_{_2}^6}{12\sqrt{35}}.
\end{eqnarray}

(c) The amplitude of $\psi(4040)$ decay into two pseudoscalar
mesons in HQET is
\begin{eqnarray}
\mathcal{M}(\psi(4040)\rightarrow
P+P)&=&\frac{2ig_{_{Iq}}g^{2}\;m_{q}\sqrt{M_{A}M_{B}M_{C}}\;\Psi(0)}
{3[(p_{_A}/2-p_{_B})^{2}-m_{q}^2]
^2}\;\epsilon_{_{A}}\cdot(p_{_B}-p_{_C}). \label{amp1}
\end{eqnarray}

(d) In  ref. \cite{yaouanc-1}, the authors gave a general
expression for calculating decays of $\psi(4040)\rightarrow
D\bar{D}$, $\psi(4040)\rightarrow D\bar{D}^{*}+\bar{D}D^{*}$ and
$\psi(4040)\rightarrow D^{*}\bar{D}^{*}$

\begin{eqnarray}
\Gamma(\psi(4040))=Ck^{3}N_{2}(k^2),
\end{eqnarray}
where $C$ means a spin-SU(3) factor corresponding to the
particular channel under consideration ($C=1/3$ for $D\bar{D}$,
$C=4/3$ for $D\bar{D}^{*}+\bar{D}D^{*}$ and $C=7/3$ for
$D^{*}\bar{D}^{*}$), $k$ is the three-momentum of the final
particles in the CM frame of $\psi(4040)$, and $N_{2}(k^2)$ is a
normalization factor and has the following expression
\begin{eqnarray}
N_{2}(k^2)=\frac{R^{3}\gamma^2
M}{43740\pi^{3/2}}[L_{2}^{3/2}(4\xi)\exp(-\xi)]^{2},
\end{eqnarray}
where $L_{2}^{3/2}$ is a Laguerre polynomial, and $\xi={k^2
R^2}/{6}$.


\begin{thebibliography}{99}
\bibitem{2317} B.Aubert et al., The BABAR Collaboration, Phys. Rev. Lett. {\bf B 90}, 242001, 2003;
F. Porter, Eur. Phys. J. {\bf C33}, S219-S222, 2004; P. Krokovny
et al., The BABAR Collaboration,
 Phys. Rev. Lett. {\bf 91}, 262002, 2003; D. Besson et al., The CLEO  Collaboration, Phys. Rev. {\bf D68}, 032002, 2003;
A. Evdokimov et al., The SLEX  Collaboration, Phys. Rev. Lett.
{\bf 93}, 242001, 2004.

\bibitem{Bardeen} W.A. Bardeen, E.J. Eichten and C.T. Hill, Phys. Rev. {\bf D 68 }, 054024, 2003;
 M.A. Nowak, M. Rho and I. Zahed, Acta Phys.Polon. {\bf B35},
 2377-2392, 2004.

\bibitem{Chao} K.D. Chao, Phys. Lett. {\bf B599}, 43-47, 2004.

\bibitem{four} T. Barnes, F. Close and H. Lipkin, Phys. Rev. {\bf D68}, 054006, 2003;
A.P. Szczepaniak, Phys. Lett. {\bf B567}, 23-26, 2003; E. Beveren
and G. Rupp, hep-ph/0305035; H.Y. Cheng and W.S. Hou, Phys. Lett.
{\bf B566}, 193-200, 2003.

\bibitem{chen} Y.Q. Chen and X.Q. Li, Phys. Rev. Lett. {\bf 93}, 232001, 2004.

\bibitem{lipkin} H. Lipkin, arXiv: {\tt hep-ph/0501209}.

\bibitem{PDG} The Data Group, Phys. Lett. {\bf B592}, 1, 2004.

\bibitem{Barnes} T. Barnes, S. Godfrey and E.S. Swanson, arXiv: {\tt
hep-ph/0505002}.

\bibitem{Micu} L. Micu, Nucl. Phys. {\bf B 10}, 521, 1969.
\bibitem{qpc} H.G. Blundell and S. Godfrey, Phys. Rev. {\bf  D53}, 3700, 1996;
A. Le Yaouanc, L. Oliver, O. P$\grave{e}$ne and J. Raynal, Phys.
Lett. {\bf B71}, 397, 1977; {\bf B72}, 57, 1977; P.R. Page, Nucl.
Phys. {\bf B446}, 189, 1995; S. Capstick and N. Isgur, Phys. Rev.
{\bf D34}, 2809, 1986.

\bibitem{yaouanc}A. Le Yaouanc, L. Oliver, O. P$\grave{e}$ne and J. Raynal,
Phys. Rev. {\bf D8}, 2223, 1973; {\bf D9}, 1415, 1974; {\bf D11},
1272, 1975; Phys. lett. {\bf B71}, 57(1977).

\bibitem{yaouanc-1}A. Le Yaouanc, L. Oliver, O. P$\grave{e}$ne and J. Raynal,
Phys. lett. {\bf B72}, 57, 1977.

\bibitem{yaouanc-book} A. Le Yaouanc, L. Oliver, O. P$\grave{e}$ne and J. Raynal,
{\it Hadron Transitions in the Quark Model}, Gordon and breach
science publishers, New York, 1987.



\bibitem{qpc-90}S. Capstick and W. Roberts, Phys. Rev.  {\bf D49}, 4570, 1994.
\bibitem{ackleh}E.S. Ackleh and T. Barnes, Phys. Rev.  {\bf D54}, 6811, 1996.
\bibitem{Zou} R.G Ping, H.Q Jiang, P.N Shen and B.S Zou, Chin. Phys. Lett. {\bf 19}, 1592, 2002;
H.Q. Zhou, R.G. Ping and B.S. Zou, Phys. Lett. {\bf B611}, 123, 2005.

\bibitem{polosa}A.D. Polosa, Riv. Nuovo Cim. {\bf 23N11}, 1, 2000.

\bibitem{Hill}W.A. Bardeen and C.T. Hill, Phys. Rev. {\bf D49},
409, 1994.

\bibitem{HQET}M. Neubert, Phys. Rep. {\bf 245}, 259 (1994).

\bibitem{falk}A.F. Falk and T. Mehen, Phys.Rev. D {\bf53}, 231,
1996.

\bibitem{hadronization}J.H. K$\ddot{u}$hn, Nucl. Phys. {\bf B157}, 125,
1979. L. Bergstr$\ddot{o}$m and P. Ernstr$\ddot{o}$m, Phys. Lett.
{\bf B267}, 111, 1991.

\bibitem{psi(4040)}R.V. Royen and V.F. Weisskopf, Nuov. Cim. {\bf
50}, 617, 1967; {\bf 51}, 583 (1967).

\bibitem{BES} BES Collaboration, Phys. Rev.  {\bf D62}, 012002, 2000.

\bibitem{Zhu}Wei Wei, Peng-Zhi Huang and Shi-Lin Zhu, arXiv: {\tt
hep-ph/0510039}.
\end{thebibliography}
\end{document}